\documentclass[%
 reprint,
 amsmath,amssymb,
 aps,
]{revtex4-1}
\usepackage{afterpage}
\usepackage{color}
\usepackage{graphicx}
\usepackage{dcolumn}
\usepackage{bm}

\newcommand{\comment}[1]{}

\begin{document}

\preprint{APS/123-QED}

\title{Signatures of Topological Phase Transition in 3d Topological Insulators from Dynamical Axion Response}

\author{Imam Makhfudz$^{\psi}$}
\affiliation{%
$^{\psi}$Laboratoire de Physique Th\'{e}orique--IRSAMC, CNRS and Universit\'{e} de Toulouse, UPS, F-31062 Toulouse, France\\ 
}%
\date{\today}

\begin{abstract}
Axion electrodynamics, first proposed in the context of particle physics, manifests itself in condensed matter physics in the topological field theory description of
3d topological insulators and gives rise to magnetoelectric effect, where applying magnetic (electric) field $\mathbf{B}(\mathbf{E})$ induces polarization (magnetization) $\mathbf{p}(\mathbf{m})$.
We use linear response theory to study the associated topological current using the Fu-Kane-Mele model of 3d topological insulators in the presence of time-dependent uniform weak magnetic field.
By computing the dynamical current susceptibility $\chi^{\mathbf{j}_p\mathbf{j}_p}_{ij}(\omega)$, we discover  from its static limit
an `order parameter' of the topological phase transition between weak topological (or ordinary) insulator and strong topological insulator, found to be continuous.The $\chi^{\mathbf{j}_p\mathbf{j}_p}_{ij}(\omega)$ shows a sign-changing singularity at a critical frequency with suppressed strength in the topological insulating state.
Our results can be verified in current noise experiment on 3d TI candidate materials for the detection of such topological phase transition.




\end{abstract}

\pacs{Valid PACS appear here}
\maketitle


\section{Introduction}
The theoretical \cite{K-M}\cite{BernevigZHang} and experimental \cite{ZHasanKaneRMP} discoveries of quantum spin Hall effect in 2d followed by its extension 
to the idea of topological insulators in 3d \cite{FKM} have become one of the most exciting developments in condensed matter physics in the $21^{\mathrm{st}}$ century \cite{QiZhangRMP}.Topological insulators (TIs) are characterized by novel response properties the most striking of which is an electromagnetic response called magnetoelectric (TME) effect in 3d topological insulators, where an applied magnetic field $\mathbf{B}$ induces electric (charge) polarization $\mathbf{p}$ (or applied electric field $\mathbf{E}$ induces magnetization $\mathbf{m}$).In topological field theory \cite{QiHughesZhang}, this effect is derived from the effective action
$S_{\theta}=\frac{ e^2}{4\pi^2 \hbar c}\int dt d^3x \theta\mathbf{E}\cdot \mathbf{B}$
known as axion electrodynamics \cite{WilczekAXION}, giving rise to magnetoelectric response $\mathbf{p}=\frac{e^2}{4\pi^2\hbar c}\theta\mathbf{B},\mathbf{m}=\frac{ e^2}{4\pi^2\hbar c}\theta\mathbf{E}$.Axion term also leads to Witten effect where a unit magnetic monopole binds a generally fractional electric charge in a medium with $\theta\neq 0$, as also studied in topological insulator context \cite{RosenbergFranz}.The $\theta$ is the pseudoscalar axion field; odd under parity and also odd under time reversal.For static time-reversal invariant systems, $\theta=0$ or $\pi$, corresponding to ordinary and topological insulators respectively.When time reversal symmetry is broken, $0<\theta<\pi$. When the system is dynamical \cite{dynamicalaxion}, the $\theta$ becomes a time-dependent function, which we will later denote by $a$. We will study this dynamical case and focus on $\mathbf{p}\sim\mathbf{B}$ response.

To study this magnetoelectric response of 3d topological insulators, we will consider Fu-Kane-Mele (FKM) model of 3d TI \cite{FKM} given by

\begin{equation}\label{3dTIFKMmodel}
 H_0=t\sum_{\langle ij\rangle}c^{\dag}_ic_j+i\frac{8\lambda_{SO}}{a^2_l}\sum_{\langle\langle ij\rangle\rangle}c^{\dag}_i\mathbf{s}\cdot\left(\mathbf{d}^1_{ij}\times\mathbf{d}^2_{ij}\right)c_j
\end{equation}
where $i,j$ are the sites of 3d diamond lattice, $c^{\dag}_i(c_i)$ is the electron creation (annihilation) operator at site $i$, $t$ the hopping integral, $\lambda_{\mathrm{SO}}$ the spin-orbit 
coupling strength, $a_l$ the cubic unit cell size, $\mathbf{s}$ the Pauli spin matrix vector, $\mathbf{d}^1_{ij}$ and $\mathbf{d}^2_{ij}$ are two consecutive nearest-neighbor vectors connecting two second-neighbor sites.The 3d topological insulators are characterized by a set of four $Z_2$ topological invariants $(\nu_0,\nu_1,\nu_2,\nu_3)$ \cite{Balents-Moore-RahulRoy}\cite{Balents-Moore-RahulRoy2}.The phase diagram of FKM model of 3d TIs contains weak topological insulator (WTI) phase ($\nu_0=0$, essentially equivalent to ordinary insulator) which is not robust to weak perturbation and disorder 
and strong topological insulator (STI) phase ($\nu_0=1$), which is robust to weak perturbation and disorder.A number of materials have been predicted to be 3d STIs, such as $\mathrm{Bi_{1-x}Sb_x}$, $\alpha-\mathrm{Sn}$, and $\mathrm{HgTe}$ under axial strain \cite{Fu-KaneTIinversionsymmetry}.The topological phase transition between the two phases is normally marked by a gap closing point \cite{MurakamiNJP}.However, the gap itself cannot be used as an `order parameter' of the transition as the gap is nonzero in both
ordinary and topological insulating phases of 3d TIs.This is what motivates our work here where we consider dynamical axion response of 3d TIs and obtain as our main results; 1)TME effect can be described in linear response theory with a vector potential involving axion field.2)Correlation function of topological current is used to define a dynamical current susceptibility which in its static limit surprisingly gives an `order parameter' of the topological phase transition.3)The (real part of) dynamical current susceptibility shows a sign-changing singularity whose strength is suppressed in the topological insulating state.The latter two results provide the signatures of the topological phase transition in 3d TIs.

\section{Current-based Formalism}Most of the existing studies on the electromagnetic response in solids employed the abstract Berry phase formalism \cite{KingSmithVanderbilt}\cite{OrtizMartin}\cite{RestaReview}.In this paper, we employ current density formalism \cite{Strinati}\cite{PRthesis} which is more physically motivated as it works directly with such measurable quantities as current.We study the response of a charged particle system to electromagnetic fields described by 4-vector potential $A_{\mu}=(A_0,\mathbf{A})$.We focus on the case with an applied magnetic field and choose a gauge where $A_0=0$\cite{orbital}.The Hamiltonian is given by 
\begin{equation}
 \hat{H}'=\int d\mathbf{x} \frac{1}{2m}|(-i\nabla -\frac{e\mathbf{A}(\mathbf{x},t)}{c})\hat{\Psi}(\mathbf{x})|^2
 \end{equation}
where we use a non-relativistic fermion system of mass $m$ and charge $e$ as an example, giving rise to charge current operator $\hat{\mathbf{j}}$
\[
\hat{\mathbf{j}}(\mathbf{x},t)=\frac{e\hbar}{2imc}\left[\hat{\Psi}^{\dag}(\mathbf{x})(\nabla\hat{\Psi}(\mathbf{x}))-(\nabla\hat{\Psi}^{\dag}(\mathbf{x}))\hat{\Psi}(\mathbf{x})\right]
 \]
 \begin{equation}
 -\frac{e^2}{mc^2}\mathbf{A}(\mathbf{x},t)\hat{\Psi}^{\dag}(\mathbf{x})\hat{\Psi}(\mathbf{x})
 \end{equation}
The first and the second terms are known as the paramagnetic current $\hat{\mathbf{j}}_p(\mathbf{x},t)$ and the diamagnetic $\hat{\mathbf{j}}_p(\mathbf{x},t)$ current, respectively;  $\hat{\mathbf{j}}(\mathbf{x},t)=\hat{\mathbf{j}}_p(\mathbf{x},t)+\hat{\mathbf{j}}_d(\mathbf{x},t)$.Such decomposition also works for relativistic fermion systems \cite{PRthesis}\cite{orbitaldirac}.
In 3d TIs, the applied magnetic field is expected to induce polarization $\mathbf{p}$.We can express the polarization in terms of the value of the current $\mathbf{j}$
\begin{equation}\label{P-j}
p_i(\omega)=\frac{i}{\omega}\int d\mathbf{r} j_i(\mathbf{r},\omega) 
\end{equation}
Within linear response theory, the induced current in response to vector potential (in a gauge where $A_0=0$) can be written in normalized units as \cite{Strinati}\cite{PRthesis}\cite{PRprl}
\begin{equation}\label{current}
\delta j_i(\mathbf{r},\omega)=\int d\mathbf{r}'\chi^{\mathbf{j}_p\mathbf{j}_p}_{ij}(\mathbf{r},\mathbf{r}',\omega)A_j(\mathbf{r}',\omega)+\rho_0(\mathbf{r})A_i(\mathbf{r},\omega)
\end{equation}
where $\chi^{\mathbf{j}_p\mathbf{j}_p}_{ij}(\mathbf{r},\mathbf{r}',\omega)$ is the paramagnetic-paramagnetic current-current correlation function, $\rho_0(\mathbf{r})$ is the ground-state density, and $\mathbf{A}(\mathbf{r},\omega)$ is the vector potential.The first and second terms in Eq. (\ref{current}) represent the paramagnetic and diamagnetic contributions respectively to the induced current.Using this current-based formalism and linear response theory to study FKM model of inversion-symmetric 3d TIs, 
we found that TME effect cannot arise from purely Maxwellian electrodynamics \cite{SuppMat}.A novel form for vector potential $\mathbf{A}$ in accordance with 
axion electrodynamics is required in order to consistently describe the TME effect.We will derive this new form of vector potential in the following section.

\section{Axion Electrodynamics}
We find the expression for $\mathbf{A}$ by solving the field equations derived from the Lagrangian involving the axion term.In the presence of axion term mentioned in the beginning, rewritten as
\begin{equation}
\Delta \mathcal{L}=\kappa a \mathbf{E}\cdot\mathbf{B} 
\end{equation}
where $\kappa$ is the coupling constant, and $a$ is the generally time-dependent axion field added to the standard Maxwellian electrodynamics described by $\mathcal{L}=(\mathbf{E}^2-\mathbf{B}^2)/2-(\rho\phi-\mathbf{j}\cdot\mathbf{A})$, the field equations are modified and are given as \cite{WilczekAXION}\cite{kineticmassaxion}
\begin{equation}\label{MAE1}
 \nabla\cdot\mathbf{E}=\rho -\kappa \nabla a \cdot \mathbf{B},
  \nabla\times\mathbf{E}=-\frac{\partial\mathbf{B}}{\partial t}, 
 \end{equation}
 \begin{equation}\label{MAE2}
 \nabla\cdot \mathbf{B}=0,\nabla\times\mathbf{B}=\frac{\partial\mathbf{E}}{\partial t} + \mathbf{j}+\kappa(\dot{a}\mathbf{B}+\nabla a\times \mathbf{E})
\end{equation}
We note the presence of extra terms in the form of additional charge density in Eq.(\ref{MAE1}) and current density in Eq.(\ref{MAE2}) due to the   axion field $a$.Considering a situation with no free charge density ($\rho$) and no free current density ($\mathbf{j}$) \textit{within} a periodic solid subjected to 
an \textit{external} spatially uniform magnetic field $\mathbf{B}$, we found that the axion field $a$ \cite{Notesinderivation} generates an electric field $\mathbf{E}$
given by
\begin{equation}\label{newE}
\mathbf{E}=\mathbf{E}_0-\kappa \mathbf{B} a 
\end{equation}
This electric field then induces polarization $\mathbf{p}$ in the solid, manifesting the TME effect. 

Since the axion field $a$ is time-dependent, the resulting electric field $\mathbf{E}$ and polarization $\mathbf{p}$
are also time-dependent.As a result, one obtains a charge current following from $\mathbf{j}=\partial_t\mathbf{p}$.Dynamical axion ensures the presence of a charge current even when the magnetic field is static.This current further modifies the vector potential $\mathbf{A}$.We found that the vector potential $\mathbf{A}$ for a general time-dependent uniform magnetic field becomes \cite{Notesassumption}
\begin{equation}\label{totalA}
\mathbf{A}(\mathbf{r},t)=\frac{\mathbf{B}(t) \times \mathbf{r}}{2} + \kappa  \int^t_{-\infty} \mathbf{B}(t') a(t') dt' 
\end{equation}
\[
 =\mathbf{A}^{\mathrm{Maxwell}}(\mathbf{r},t)+\mathbf{A}^{\mathrm{axion}}(\mathbf{r},t)
\]
We note that $\mathbf{A}$ now acquires a new term in addition to the standard $\mathbf{A}^{\mathrm{Maxwell}}(\mathbf{r},t)=(\mathbf{B}(t)\times\mathbf{r})/2$; 
the axion term $\mathbf{A}^{\mathrm{axion}}(\mathbf{r},t)=\kappa  \int^t_{-\infty} \mathbf{B}(t') a(t') dt' $, which is proportional to the applied uniform magnetic field $\mathbf{B}$, 
implying that the polarization $\mathbf{p}$ acquires a component along the direction of $\mathbf{B}$, according to Eqs.(\ref{P-j}) and (\ref{current}).This is the topological polarization of the TME effect. 

\section{Dynamical Axion Response}
We now consider inversion-symmetric Fu-Kane-mele model of 3d TIs \cite{FKM} for concreteness, where the Maxwellian
$(\mathbf{B}(t)\times\mathbf{r})/2$ term generates no polarization \cite{SuppMat}, so that we can focus on the topological part of $\mathbf{A}$ of interest.Considering general space-time dependent axion field $a(\mathbf{r},t)$ and treating the paramagnetic and diamagnetic parts on equal footing \cite{PRprl},
we obtain the following equation for the topological induced current,
\[
\delta j_i(\mathbf{r},\omega)=
\]
\begin{equation}\label{totalcurrentomega}
\int d\mathbf{r}'\left[ \chi^{\mathbf{j}_p\mathbf{j}_p}_{ij}(\mathbf{r},\mathbf{r}',\omega)A^{\mathrm{axion}}_j(\mathbf{r}',\omega)-\chi^{\mathbf{j}_p\mathbf{j}_p}_{ij}(\mathbf{r},\mathbf{r}',0)A^{\mathrm{axion}}_j(\mathbf{r},\omega)\right]
\end{equation}
where the paramagnetic and diamagnetic contributions take parallel form.We seek to characterize the response of topological insulators in terms of this quantity. 

Consider the (retarded) total current-total current correlation function, directly connected to linear response theory, given by
\begin{equation}\label{retardedcorrfunction}
\chi^{\mathbf{j}\mathbf{j}}_{ij}(\mathbf{r},\mathbf{r}',\omega)=-\frac{i}{\hbar}\int^0_{-\infty}d\tau e^{i\omega\tau}\langle G | \left[\hat{j}^I_{i}(\mathbf{r},0),\hat{j}^I_{j}(\mathbf{r}',\tau)\right]|G\rangle
\end{equation}
where $\langle G | \cdots | G \rangle$ represents the ground state expectation value,  $\hat{\mathbf{j}}^I_{}(\mathbf{x},\tau)=\exp(iH_0\tau/\hbar)\hat{\mathbf{j}}(\mathbf{x},\tau)\exp(-iH_0\tau/\hbar)$ the current operator in interaction picture \cite{SuppMat}.Now we define the dynamical current susceptibility
\begin{equation}
\chi^{\mathbf{j}\mathbf{j}}_{ij}(\omega)= \int d\mathbf{r}\int d\mathbf{r}'\chi^{\mathbf{j}\mathbf{j}}_{ij}(\mathbf{r},\mathbf{r}',\omega) 
\end{equation}
where $ \int d\mathbf{r},\int d\mathbf{r}'$ are integrals over the unit cell of the lattice (with unit volume in our unit system).Using Eqs.(\ref{current}),(\ref{totalcurrentomega}), and (\ref{retardedcorrfunction}), exact derivation \cite{SuppMat} gives 
\begin{equation}\label{totaldyncurrsuscept}
\chi^{\mathbf{j}\mathbf{j}}_{ij}(\omega)=\chi^{\mathbf{j}_p\mathbf{j}_p}_{ij}(\omega)+N_s\delta_{ij}
\end{equation}
where 
\begin{equation}\label{etafunction}
\chi^{\mathbf{j}_p\mathbf{j}_p}_{ij}(\omega)= \int d\mathbf{r}\int d\mathbf{r}'\chi^{\mathbf{j}_p\mathbf{j}_p}_{ij}(\mathbf{r},\mathbf{r}',\omega) 
\end{equation}
with $\chi^{\mathbf{j}_p\mathbf{j}_p}_{ij}(\mathbf{r},\mathbf{r}',\omega)$ is given by Eq.(\ref{retardedcorrfunction}) but with $\hat{\mathbf{j}}_p(\mathbf{x},\tau)$ in place of  $\hat{\mathbf{j}}(\mathbf{x},\tau)$ and $N_s=\int d\mathbf{x}\langle G |\hat{\Psi}^{\dag}(\mathbf{x})\hat{\Psi}(\mathbf{x})|G\rangle $ is the ground state average number of electrons per unit cell for each spin species, which is a given number.The second term in Eq.(\ref{totaldyncurrsuscept}) thus merely acts as a constant shift to the first term. 

The result Eq.(\ref{totaldyncurrsuscept}) suggests that we can adopt $\chi^{\mathbf{j}_p\mathbf{j}_p}_{ij}(\omega)$ to study the response of 3d TIs and its dependence on the frequency and microscopic parameters of $H_0$, especially the hopping integral, as we tune the system across its topological phase transition.This $\chi^{\mathbf{j}_p\mathbf{j}_p}_{ij}(\omega)$ can be used as a dynamical characterization of the TME effect of 3d TI in response to a uniform magnetic field applied along an appropriate direction.The response will be dominated by the diagonal elements of $\chi^{\mathbf{j}_p\mathbf{j}_p}_{ij}(\omega)$ because in TME effect $\mathbf{p}$ is collinear with $\mathbf{B}$.We will explicitly compute $\chi^{\mathbf{j}_p\mathbf{j}_p}_{zz}(\omega)$ as example.
This response function Eq.(\ref{etafunction}) is determined entirely by the intrinsic band structure properties of the insulators.Physically, $\chi^{\mathbf{j}\mathbf{j}}_{ij}(\omega)$($\chi^{\mathbf{j}_p\mathbf{j}_p}_{ij}(\omega)$) represents the correlation between the total currents (induced topological currents) at two different points, averaged over space, as it is related to 
two-particle Green's function \cite{Strinati}.This quantity therefore encodes and should be able to detect the nonlocal global topological properties of the band structure 
that manifests in the electromagnetic response of the insulator. 

We calculated the function $\chi^{\mathbf{j}_p\mathbf{j}_p}_{zz}(\omega)$ using the general expression for $\chi^{\mathbf{j}_p\mathbf{j}_p}_{ij}(\mathbf{r},\mathbf{r}',\omega)$ \cite{PRthesis} 
derived from two-particle Green's function formalism \cite{Strinati},
\[
\chi^{\mathbf{j}^p\mathbf{j}^p}_{\alpha\beta}(\mathbf{r},\mathbf{r}',\omega)=-\frac{i}{\hbar}\int^0_{-\infty}d\tau e^{i\omega\tau}\langle G | \left[\hat{j}^{I p}_{\alpha}(\mathbf{r},0),\hat{j}^{I p}_{\beta}(\mathbf{r}',\tau)\right]|G\rangle
\]
\begin{equation}\label{currentsuscept}
=\sum_{\mathbf{k},i,f}\frac{g^{(\mathbf{k})}_{if}}{N_{\mathbf{k}}}
	\left[	\frac{(\psi^*_{i\mathbf{k}}(\mathbf{r})\hat{j}^p_{\alpha}\psi_{f\mathbf{k}}(\mathbf{r}))(\psi^*_{f\mathbf{k}}(\mathbf{r}')\hat{j}^p_{\beta}\psi_{i\mathbf{k}}(\mathbf{r}'))}{\epsilon_{i\mathbf{k}}-\epsilon_{f\mathbf{k}}+\omega+i\eta}+c.c.|_{-\omega}\right]
		\end{equation}
	where $N_{\mathbf{k}}$ is the number of unit cells (the number of $\mathbf{k}$ points in the first Brillouin zone), $g^{(\mathbf{k})}_{if}=(f_{i\mathbf{k}}-f_{f\mathbf{k}})/(1+\delta_{if})$, $f_{i\mathbf{k}}$ being the Fermi-Dirac distribution function at $T=0$ of the $i^{\mathrm{th}}$ Bloch state $\psi_{i\mathbf{k}}(\mathbf{r})$ at wavevector $\mathbf{k}$ corresponding to the $i^{\mathrm{th}}$ energy band $\epsilon_{i\mathbf{k}}$.The paramagnetic current operator is $\hat{\mathbf{j}}^p_{}=-i\left(\nabla-\nabla^{\dag}\right)/2$ where
	$\nabla$ acts on everything to its right while $\nabla^{\dag}$ acts on every thing to its left.The $i$ and $f$ denote the initial and final states, taken to be the ground state and excited states respectively.In this work, the $g^{(\mathbf{k})}_{i,f}$ simply equals 0
	for $i=f$ and 1 for $i\neq f$.  We computed the Bloch states from the FKM model Eq.(\ref{3dTIFKMmodel}).Our calculation applies to 3d TIs described by $H_0$ in Eq.(\ref{3dTIFKMmodel}) with  $0< N_s\leq 1$.

\section{Results:\space Signatures of Topological Phase Transition.}We show the profile for $\chi^{\mathbf{j}_p\mathbf{j}_p}_{zz}(\omega)$ in the static limit $\chi^{\mathbf{j}_p\mathbf{j}_p}_{zz}(0)$ 
corresponding to static susceptibility in Fig.1,
demonstrating its variation as function of the deformation $dt$ of hopping integral $t=1.0+dt$ along two different directions; $[111]$ and $[1,-1,-1]$ directions, each separately and simultaneously.The deformation of hopping integral can be accomplished by appropriate chemical doping which creates point defects of the type vacancies or substitutional defects in 
the semiconductors \cite{Chemistry}.The topological phase diagram obtained in Fu-Kane-Mele's seminal work \cite{FKM} suggests that one has WTI (reducible to ordinary insulator) for $dt<0$, STI (true topological insulator) for $dt>0$,  
and quantum critical point semimetal at $dt=0$ for deformation along [111] while keeping $dt=0$ along $[1,-1,-1]$ and vice versa, as well as when the hopping integral is deformed along the two directions simultaneously. 

Fig.1 interestingly suggests that the two topologically distinct insulating phases are distinguished by different behaviors 
of our static susceptibility in the following aspects:1.The curves corresponding to separate hopping deformations are linear functions with distinct slopes.The deflection occurs right at the quantum critical point semimetal.2.The simultaneous deformation gives rise to a close to the average of the static current susceptibilities of the separate deformations in the WTI phase whereas 
in the STI phase, the former shows clear deviation away from the other two.These changes in behavior can be taken as an indication of the topological phase transition.
\begin{figure}
 \includegraphics[scale=0.60]{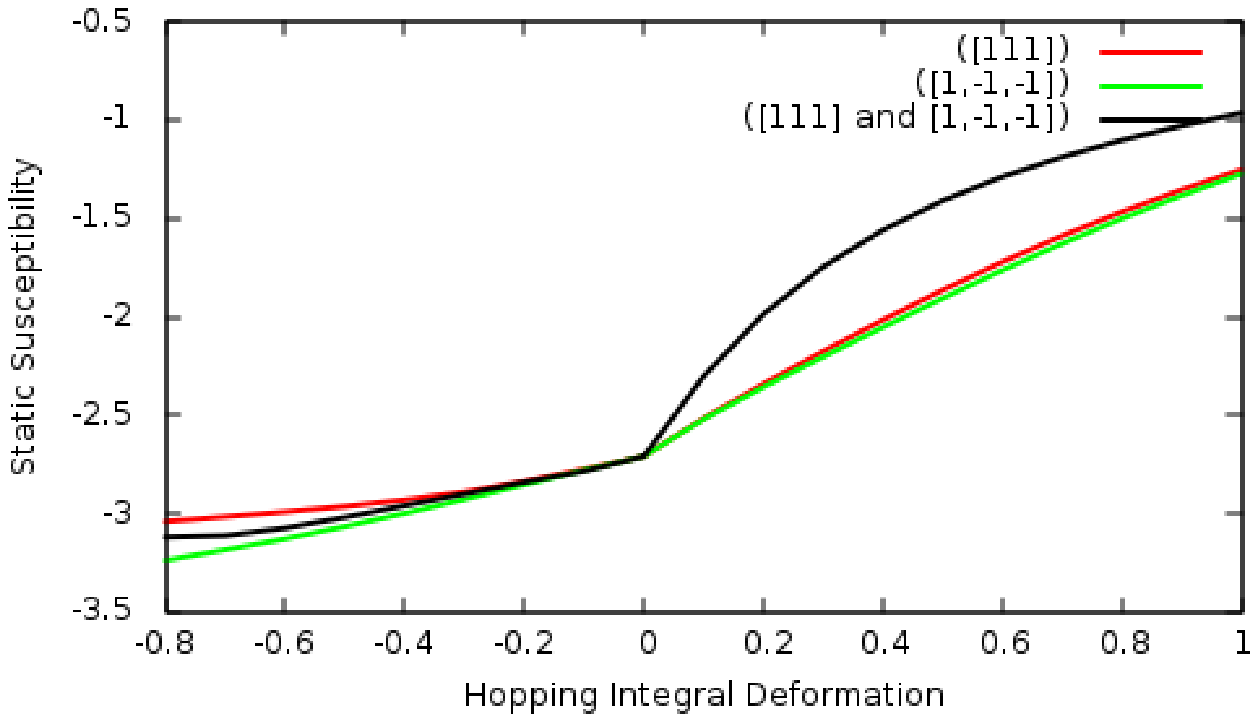}
 \includegraphics[scale=0.60]{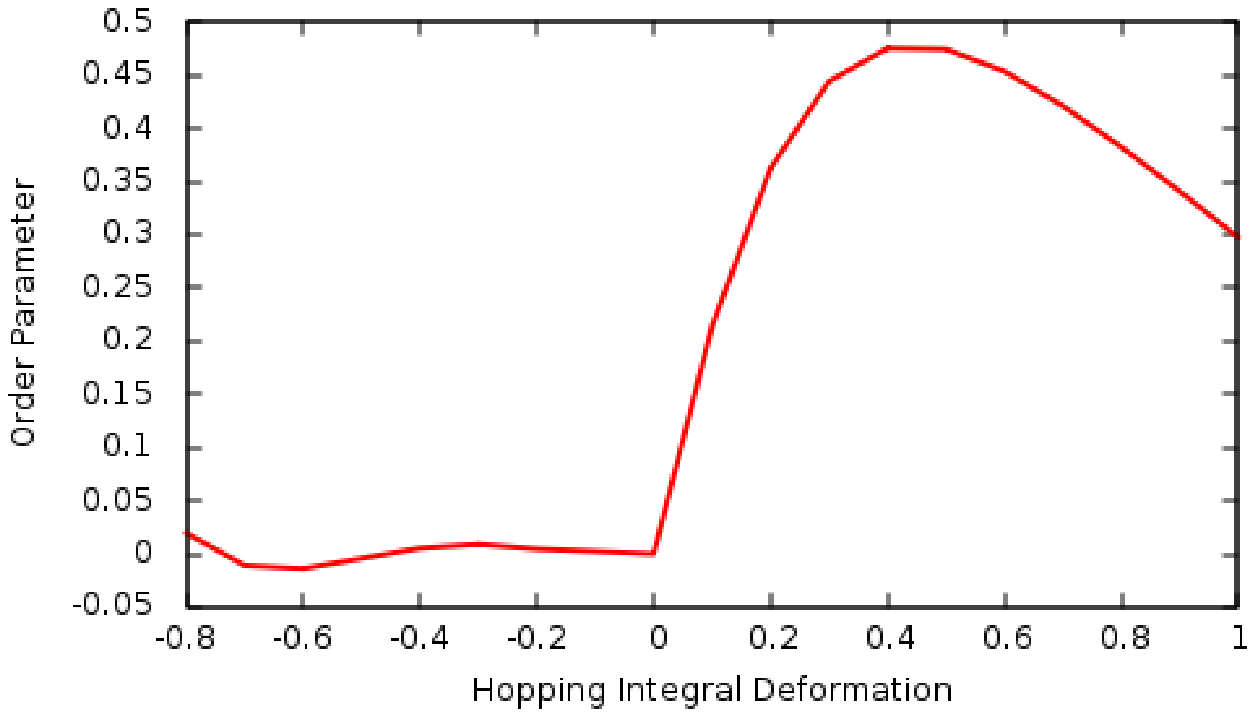}
 \label{fig:PvsGap}
 \caption{a)The (real part of) static susceptibility $\chi^{\mathbf{j}_p\mathbf{j}_p}_{zz}(\omega=0)$ \cite{Noteimag} as function of hopping integral deformation $dt$ (proportional to band gap $\Delta=2|dt|$) 
 along directions $[111]$,$[1,-1,-1]$, and both, with equal $dt$. 
 b) The resulting `order parameter' $\delta \chi^s (dt)$ of the topological phase transition $\delta p_0(dt)$ given in Eq.(\ref{OrderParameter}).The units are such that $t=1.0$ and $a_l=1.0$ and we have set $\lambda_{SO}=0.125$.}
\end{figure}
The above observations suggest that we can define a quantity as function of deformation $dt$ which can serve as an `order parameter' of the topological phase transition;
\begin{equation}\label{OrderParameter}
 \delta \chi^s (dt) = \chi^{\mathbf{j}_p\mathbf{j}_p}_{zz}(0)_{[111]+[1,-1,-1]}(dt)-\overline{\chi^{\mathbf{j}_p\mathbf{j}_p}_{zz}(0)(dt)}
 \end{equation}
 where
\begin{equation}
 \overline{\chi^{\mathbf{j}_p\mathbf{j}_p}_{zz}(0)(dt)}=\frac{\chi^{\mathbf{j}_p\mathbf{j}_p}_{zz}(0)_{[111]}(dt)+\chi^{\mathbf{j}_p\mathbf{j}_p}_{zz}(0)_{[1,-1,-1]}(dt)}{2}
\end{equation}
Fig.1 shows that $\delta \chi^s (dt)$ goes from zero to nonzero across the topological phase transition and can therefore be used as an unambiguous signature of the occurrence of such transition.Combining Eqs.(\ref{totaldyncurrsuscept}) and (\ref{OrderParameter}), it is clear that the total current susceptibility $\chi^{\mathbf{j}\mathbf{j}}_{zz}(\omega=0)$ measurable in experiment will also give the same critical behavior.

This critical behavior of $\delta \chi^s (dt)$ is nontrivial because naively, the susceptibility should only depend on the projection of the direction of electron hopping
to the direction of applied magnetic field and thus should not depend on how the deformation of the hopping integral is implemented, given the two fixed directions.This turns out to hold only if the solid is always in the same phase relevant to its current response, e.g. weak topological (equivalent to ordinary) insulating phase.Now the current susceptibility shows such unexpected behavior as one crosses a topological phase transition, even if the two directions of hopping deformation are fixed. $\delta \chi^s (dt)$ thus can serve as an `order parameter' for this topological phase transition which Fig.1 suggests to be continuous, in agreement with experiments \cite{Exps}.This is also consistent with the expectation for a phase transition obtained by continuously tuning a parameter in a non-interacting fermion model Eq.(\ref{3dTIFKMmodel}).The computed static orbital polarizability also agrees with the result of Berry phase theories \cite{Vanderbilt}.

Theoretically, this nontrivial behavior originates from the $C_2$ symmetry of the band structure of the FKM model Eq.(\ref{3dTIFKMmodel}) under $\pi$ rotation about the $x$ axis 
preserved under simultaneous deformation of the hopping integrals with equal strength along $[111]$ and $[1,-1,-1]$ directions, but broken when the deformation is applied separately 
in one of the two directions.This $C_2$ symmetry manifests its $Z_2$ effect $\sim (-1)^{\nu_0}$ in the current susceptibility for (strong) topological insulator since 
the surface Fermi arc encloses odd number of Dirac points ($\nu_0=1$)
while the effect vanishes in the weak topological (ordinary) insulator since the arc encloses even number of them ($\nu_0=0$)\cite{FKM}.We have checked and confirmed numerically that such a critical behavior in $ \delta \chi^s (dt)$ would never occur if we set the spin-orbit coupling to zero for example, where only one type of insulator; ordinary insulator (at any $0 < |dt| \leq t$) could exist \cite{SuppMat}. 

We now present the outcome for the dynamical response function $\chi^{\mathbf{j}_p\mathbf{j}_p}_{zz}(\omega)$ in the ordinary and topological insulator phases as well as at the quantum critical point between them in Fig.2.Apart from  the numerical fluctuations due to the finite-size effect around some critical frequencies, the profile of $\chi^{\mathbf{j}_p\mathbf{j}_p}_{zz}(\omega)$ is consistent with the familiar $\frac{1}{\omega-\omega_0}$ behavior of the two-body (e.g.density-density, current-current) response functions that have a sign-changing singularity
at a critical frequency $\omega_0$ at which the functions flip sign.Using Eq.(\ref{totaldyncurrsuscept}) and noting the magnitude of $\chi^{\mathbf{j}_p\mathbf{j}_p}_{zz}(0)$ from Figs.1 and 2, it is clear that the singularity and the sign change also occur to the $\chi^{\mathbf{j}\mathbf{j}}_{zz}(\omega)$ measurable in experiment.The ordinary and topological insulator phases can be distinguished by the overall strength of the singularity around $\omega_0$.In general, we find that topological insulating phase leads to 
suppression of the strength of singularity  in the dynamical response function at around $\omega_0$.The switch from negative to positive correlation across $\omega_0$ is nontrivial 
but can be nicely explained in terms of the continuity equation obeyed by the charge density and the current induced by the polarization generated by the magnetic field via the TME effect \cite{SuppMat}.

\begin{figure}
 \includegraphics[scale=0.25]{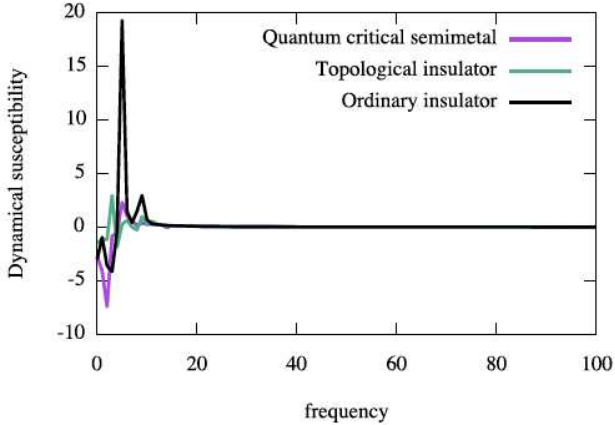}
  \label{fig:DynRespSTI}
 \caption{The (real part of) response function $\chi^{\mathbf{j}_p\mathbf{j}_p}_{zz}(\omega)$ \cite{Noteimag}  for STI (topological insulator), WTI (or ordinary insulator), and the critical point (semimetal) as function of frequency $\omega$.We have used $dt=\pm 0.80$ along direction $[1,1,1]$ for the topological (ordinary) insulator phase as example.The units are such that $t=1.0$ and $a_l=1.0$ and we have set $\lambda_{SO}=0.125$.}
\end{figure}

\section{Discussion}
Topological phase transition occurs between different phases which cannot be distinguished by a local order parameter with true symmetry breaking in the sense of Ginzburg-Landau theory \cite{Wenbook}.Rather,they are described by a global topological property.The quantum Hall transition \cite{QHtransition} 
is one of the first examples of such non-Ginzburg-Landau type of phase transition.The quantum spin Hall effect and topological insulators also display topological phase transition between
ordinary insulator and quantum spin Hall (topological insulator) phases, marked by a gap closing in the bulk excitation spectrum.Our work discovered
an `order parameter' derived from the nonlocal two-particle Green's function to probe this latter phase transition between the states distinguished by the global topological properties.

The response function that we define is determined by the current-current correlation function which amounts to a current noise power measurement. Here we propose the experimental setup for such measurement on 3d TI candidate materials described by FKM model \cite{Fu-KaneTIinversionsymmetry} to detect such topological phase transition.Consider a thick slab sample of such 3d TI materials having front and rear surfaces normal to the $[111]$ or $[1,-1,-1]$ direction but parallel to a uniform magnetic field  and are coated with insulating ferromagnet \cite{SuppMat}.The magnon of the ferromagnet provides the dynamical axion field that couples to the magnetic field and the induced electric field \cite{QiHughesZhang}. 

Time-dependent uniform magnetic field induces current 
on the two surfaces, assumed uniformly distributed, and we propose to measure their correlation. 
The current noise cross-corrrelation is given by \cite{Imry} 
\begin{equation}
 S^I_{ij}(\omega)=\frac{1}{2\pi}\int^{\infty}_{-\infty}\langle \Delta I_i(0)\Delta I_j(t)\rangle e^{-i\omega t} dt
\end{equation}
where the current noise is $\Delta I_i(t)=I_i(t)-\overline{I}_i$ at the $i=1,2$-th surface with $\overline{I}_i$ the corresponding steady-state equilibrium current (without the applied magnetic field).
The surface current will induce a time-dependent change in the charge density at the top and bottom ends of the sample, satisfying continuity equation so as to ensure charge conservation.Our theory predicts that the behavior of $S^I_{12}(\omega)$ is given by the profiles of the response function $\chi^{\mathbf{j}_p\mathbf{j}_p}_{zz}(\omega)$ shown in Figs.1 and 2.We have thus demonstrated a characterization of topological phase transition in 3d topological insulators from their dynamical axion response properties.It is important to note that our results based on current-based formalism are consistent with those obtained from Berry phase formalism, as both approaches manifest the direct correspondence between the band topology and electromagnetic response properties that characterize topological insulators.Our results are also universal as they hold for any 3d tight-binding fermion systems that host topological and ordinary insulating phases, including ultracold atom system realizing the axion electrodynamics \cite{axioncoldatom}.
 
\section{Acknowledgements}The author was supported by the grant No.ANR-10-LABX-0037 of the Programme des Investissements d'Avenir of France.The author thanks Dr.P.Romaniello and Dr.A.Berger for the guidance all along and Prof.I.Affleck for the insightful discussion and comments.

\section{Appendices}

\appendix

In these Appendices, we first  prove the absence of topological magnetoelectric effect in purely Maxwellian electrodynamics. 
We next discuss the linear response theory and the associated current-current correlation function and derive the Eq.(\ref{totaldyncurrsuscept}) in the main text.
We then show a consistency check of our results.
Finally, we give the details of the experimental setup of the current noise correlation measurement and explain how our results agree with simple physical intuition.

\bigskip






\section{Polarization from Maxwellian Electrodynamics in Inversion-symmetric Lattice Systems}

We will show that topological magnetoelectric effect cannot arise from purely Maxwellian electrodynamics. We illustrate this from linear response theory applied to Maxwellian electrodynamics 
to study the electromagnetic response of inversion-symmetric lattice periodic solids, where magnetic field is found to not generate any polarization. In inversion-asymmetric system, such polarization will be nonzero. 
Topological polarization coming from TME effect should be nonzero, regardless of symmetry property of the system under inversion and should therefore come from new term in the vector potential $\mathbf{A}$ in addition to 
the familiar $(\mathbf{B}\times \mathbf{r})/2$ term, as we derived in the main text. Here follows the proof.

Standard Maxwellian electrodynamics suggests that we can write the solution of the equation $\mathbf{B}=\nabla \times \mathbf{A}$ as usual, which in symmetric gauge is given by  
\begin{equation}\label{symmetricgauge}
\mathbf{A}(\mathbf{r},\omega)=\frac{\mathbf{B}(\mathbf{r},\omega)\times \mathbf{r}}{2}
\end{equation}
valid for spatially uniform magnetic field $\mathbf{B}$. For single molecule systems, it has been found that the diamagnetic and paramagnetic contributions can be treated on equal footing and yield the reference-independent expression 
for current and polarization in the $\omega\rightarrow 0$ limit \cite{PRprl}.
\begin{equation}\label{P-jsinglemolecule}
\delta p_i( 0)=\mathrm{lim}_{\omega\rightarrow 0}\frac{i}{\omega}\int d\mathbf{r} \int d\mathbf{r}'\chi^{\mathbf{j}_p\mathbf{j}_p}_{ij}(\mathbf{r},\mathbf{r}',\omega)\left[\frac{\mathbf{B}(\omega)\times (\mathbf{r}'-\mathbf{r})}{2}\right]_j 
\end{equation}
The above expression with its relative position vector term  $\mathbf{r}'-\mathbf{r}$ is also the one required for lattice periodic solid systems based on the requirement of lattice translational invariance, which is realized by the $\mathbf{r}'-\mathbf{r}$.
The current-current correlation function itself is also translationally invariant by definition.
\begin{equation}
 \chi^{\mathbf{j}_p\mathbf{j}_p}_{ij}(\mathbf{r}+\mathbf{R}_i,\mathbf{r}'+\mathbf{R}_i,\omega)=\chi^{\mathbf{j}_p\mathbf{j}_p}_{ij}(\mathbf{r},\mathbf{r}',\omega)
\end{equation}
for translations by arbitrary lattice vectors \cite{GiulianiS}. Note that the two independent position vectors $\mathbf{r}$ and $\mathbf{r}'$ are translated by the same lattice vector $\mathbf{R}_i$ in this definition of translational invariance.
For lattice-periodic solid systems, the Eq.(\ref{P-jsinglemolecule}) holds naturally due to the lattice periodicity and so we have
\begin{equation}\label{P-j0}
\delta p_i(0)=\mathrm{lim}_{\omega\rightarrow 0}\frac{i}{\omega}\epsilon_{ijk}\int d\mathbf{r} \int d\mathbf{r}'\chi^{\mathbf{j}_p\mathbf{j}_p}_{ij}(\mathbf{r},\mathbf{r}',\omega)\frac{B_k(\omega)(r'_l-r_l)}{2} 
\end{equation}

In this work, we are studying the linear response to a spatially uniform macroscopic magnetic field, taken to be in $z$ for concreteness. 
Writing the explicit form for each component of the polarization from Maxwellian vector potential given in Eq.(\ref{P-j0}), it can be easily checked that in current-based formalism, 
one can only obtain polarization parallel to the applied magnetic field from the contribution of the off-diagonal elements of the current susceptibility tensor. 
Diagonal elements of the susceptibility tensor only contribute to the polarization perpendicular to the magnetic field; polarization on $xy$ plane for magnetic field on $z$ direction.
The TME response described by the equation $\mathbf{p}=\frac{e^2}{4\pi^2\hbar c}\theta\mathbf{B}$ stated in the introduction in the main text on the other hand
implies that the topological polarization is parallel to the magnetic field. We will first show that for inversion symmetric systems, 
both the diagonal and off-diagonal elements of the tensor $\chi^{\mathbf{j}_p\mathbf{j}_p}_{i\neq j}$ give no contribution to the polarization $\mathbf{p}$ in Eq.(\ref{P-j0})
from the standard form of vector potential $\mathbf{A}=(\mathbf{B}\times \mathbf{r})/2$. 
This conclusion is applicable to inversion-symmetric 3d topological insulators, such as the FKM model of 3d topological insulators on diamond lattice \cite{FKM}\cite{Fu-KaneTIinversionsymmetry} with diamond-shaped Brillouin zone, which has inversion symmetry and results
in the two-fold degeneracy of the upper and lower energy bands. The key to the cancellation of the off-diagonal tensor element contribution to polarization is this inversion symmetry of the model.
The absence of this contribution to polarization can be shown as follows. 

We want to compute polarization from Eq.(\ref{P-j0}). Let us make the following substitutions $\mathbf{r}\rightarrow
 -\mathbf{r}, \mathbf{r}'\rightarrow -\mathbf{r}'$. As a result, we obtain
 \begin{equation}\label{P-j1}
\delta p_i(0)=\mathrm{lim}_{\omega\rightarrow 0}\frac{i}{\omega}\epsilon_{jkl}\int d\mathbf{r} \int d\mathbf{r}'\chi^{\mathbf{j}_p\mathbf{j}_p}_{ij}(-\mathbf{r},-\mathbf{r}',\omega)\frac{B_k(\omega)(r'_l-r_l)}{2} 
\end{equation}
where indices appearing twice are to be summed over.
For a system with inversion symmetry, we have $\chi^{\mathbf{j}_p\mathbf{j}_p}_{ij}(-\mathbf{r},-\mathbf{r}',\omega)=\chi^{\mathbf{j}_p\mathbf{j}_p}_{ij}(\mathbf{r},\mathbf{r}',\omega)$. Therefore, Eq.(\ref{P-j1}) can be written as
\begin{equation}\label{P-j2}
\delta p_i(0)=-\mathrm{lim}_{\omega\rightarrow 0}\frac{i}{\omega}\epsilon_{jkl}\int d\mathbf{r} \int d\mathbf{r}'\chi^{\mathbf{j}_p\mathbf{j}_p}_{ij}(\mathbf{r},\mathbf{r}',\omega)\frac{B_k(\omega)(r'_l-r_l)}{2} 
\end{equation}
Comparing Eqs.(\ref{P-j0}) and (\ref{P-j2}) we arrive at the conclusion that $\delta p_i=0$.
In summary,
the contribution of the off-diagonal elements of the susceptibility tensor $\chi^{\mathbf{j}_p\mathbf{j}_p}_{ij}$ to the polarization from the Maxwellian vector potential 
$\mathbf{A}(\mathbf{r},\omega)=(\mathbf{B}(\omega)\times\mathbf{r})/2$ in our problem is identically zero at all frequencies $\omega$.
This result does not imply that TME effect, which only requires time reversal symmetry on system with spin-orbit coupling, occurs only in inversion asymmetric 3d topological insulators. 
TME effect does exist in inversion symmetric TI's but their net TME vanishes.
For inversion asymmetric system, the familiar $\mathbf{A}=(\mathbf{B}\times \mathbf{r})/2$ contributes nonzero $\mathbf{p}$ but which does not have topological character. 
TME effect should exist in 3d TI regardless of whether the system is inversion symmetric or not and so this effect cannot arise from 
the Maxwellian solution for $\mathbf{A}$ in order for the linear response theory formulation of the effect to be consistent. The complete form of the vector potential $\mathbf{A}$
is as given in Eq.(\ref{totalA}) in the main text, consisting of the Maxwellian and axionic parts \cite{Notes}.

\section{Linear Response Theory and Current-Current Correlation Function}

In linear response theory, the linear response of the system to an external perturbation is represented by the change in the expectation value of an observable coupled to the external potential. This is described by the Kubo formula

\begin{equation}
\delta\langle \hat{O}(\mathbf{x},t)\rangle = \frac{i}{\hbar}\int^t_{-\infty} dt'\langle G |[H_I'(t'),\hat{O}^I(\mathbf{x},t)]|G\rangle
\end{equation}
where $\langle G|\cdots|G\rangle$ represents the ground state expectation value, $\hat{O}$ is the observable operator, and $H'_I$ is the external perturbation Hamiltonian, which are given by
\begin{equation}
H'_I(t')=e^{\frac{iH_0 t'}{\hbar}}H'(t')e^{-\frac{iH_0 t'}{\hbar}}
\end{equation}
\begin{equation}
 \hat{O}^I(\mathbf{x},t)=e^{\frac{iH_0 t}{\hbar}}\hat{O}(\mathbf{x},t)e^{-\frac{iH_0 t}{\hbar}}
\end{equation}
in the interaction picture, where $H_0$ is the unperturbed Hamiltonian and
\begin{equation}
H'(t')=\int d\mathbf{x}' \hat{O}(\mathbf{x}',t')f(\mathbf{x}',t')
\end{equation}
with $f(\mathbf{x}',t')$ an appropriate perturbation field.  

In our work, the observable is the total charge current $\hat{\mathbf{j}}(\mathbf{x},t)=\hat{\mathbf{j}}_p(\mathbf{x},t)+\hat{\mathbf{j}}_d(\mathbf{x},t)$ 
where the first and second terms represent the paramagnetic and diamagnetic contributions respectively given by
\begin{equation}\label{paracurrent}
\hat{\mathbf{j}}_p(\mathbf{x},t)=\frac{e\hbar}{2imc}
\left[\hat{\Psi}^{\dag}(\mathbf{x})
\left(\nabla\hat{\Psi}(\mathbf{x})\right)-\left(\nabla\hat{\Psi}^{\dag}(\mathbf{x})\right)\hat{\Psi}(\mathbf{x})\right]
\end{equation}
\begin{equation}\label{diamcurrent}
\hat{\mathbf{j}}_d(\mathbf{x},t)=-\frac{e^2}{mc^2}\mathbf{A}(\mathbf{x},t)\hat{\Psi}^{\dag}(\mathbf{x})\hat{\Psi}(\mathbf{x})
\end{equation}
The total induced current in response to vector potential is therefore given by 
\begin{equation}
\delta \mathbf{j}(\mathbf{x},t)=\delta\langle \hat{\mathbf{j}}(\mathbf{x},t)\rangle = \frac{i}{\hbar}\int^t_{-\infty} dt'\langle G |[H_I'(t'),\hat{\mathbf{j}}^I(\mathbf{x},t)]|G\rangle
\end{equation}
where the current operator in the interaction picture is given by $\hat{\mathbf{j}}^I(\mathbf{x},t)=\exp(iH_0 t/\hbar)\hat{\mathbf{j}}(\mathbf{x},t)\exp(-iH_0 t/\hbar)$. Using 
\begin{equation}
H'(t')=-\int d\mathbf{x}'\hat{\mathbf{j}}(\mathbf{x}',t')\cdot \mathbf{A}(\mathbf{x}',t')
\end{equation}
we obtain
\begin{equation}\label{totalinducedcurrent}
\delta j_i(\mathbf{x},t)= \int^{\infty}_{-\infty} dt' \int d\mathbf{x}'\chi^{\hat{\mathbf{j}}\hat{\mathbf{j}}}_{ij}(\mathbf{x},t;\mathbf{x}',t')A_j(\mathbf{x}',t')\end{equation}
where we have used Planck units where $\hbar=1,c=1$ and defined a retarded total current-total current correlation function
\begin{equation}\label{retardedcorrfunctiontotalj}
\chi^{\hat{\mathbf{j}}\hat{\mathbf{j}}}_{ij}(\mathbf{x},t;\mathbf{x}',t')=-\frac{i}{\hbar}\theta(t-t')\langle G | [j^I_i(\mathbf{x},t),j^I_j(\mathbf{x}',t')]|G\rangle
\end{equation}
Taking the time Fourier transform of this retarded correlation function, we obtain 
\[
\chi^{\mathbf{j}\mathbf{j}}_{ij}(\mathbf{x},\mathbf{x}',\omega)=\int^{\infty}_{-\infty}d(t'-t)e^{i\omega(t'-t)}\chi^{\hat{\mathbf{j}}\hat{\mathbf{j}}}_{ij}(\mathbf{x},t;\mathbf{x}',t')
\]
\begin{equation}\label{totalcurrentcorrelationS}
=-\frac{i}{\hbar}\int^0_{-\infty}d\tau e^{i\omega\tau}\langle G | \left[\hat{j}^I_{i}(\mathbf{x},0),\hat{j}^I_{j}(\mathbf{x}',\tau)\right]|G\rangle
\end{equation}
where we have defined $\tau=t'-t$, which takes the range $-\infty\leq\tau\leq 0$ due to the $\theta(t-t')$. The time Fourier transform of Eq.(\ref{totalinducedcurrent}) is
\begin{equation}\label{totalinducedcurrentFT}
\delta j_i(\mathbf{x},\omega)= \int d\mathbf{x}'\chi^{\hat{\mathbf{j}}\hat{\mathbf{j}}}_{ij}(\mathbf{x},\mathbf{x}',\omega)A_j(\mathbf{x}',\omega)
\end{equation}

It is known from standard many-body physics literature \cite{Strinati} that the total induced current in linear response theory is given by
\[
\delta j_i(\mathbf{x},t)= \int^{\infty}_{-\infty} dt' \int d\mathbf{x}'\chi^{\hat{\mathbf{j}}_p\hat{\mathbf{j}}_p}_{ij}(\mathbf{x},t;\mathbf{x}',t')A_j(\mathbf{x}',t')+\rho_0(\mathbf{x})A_i(\mathbf{x},t)
\]
\begin{equation}\label{totalinducedcurrentStrinati}
=\int^{\infty}_{-\infty} dt' \int d\mathbf{x}'\left[\chi^{\hat{\mathbf{j}}_p\hat{\mathbf{j}}_p}_{ij}(\mathbf{x},t;\mathbf{x}',t')+\rho_0(\mathbf{x}') \delta_{ij}\delta(x'^{\mu}-x^{\mu})\right]A_j(\mathbf{x}',t')
\end{equation}
where $\delta(x'^{\mu}-x^{\mu})=\delta(\mathbf{x}'-\mathbf{x})\delta(t'-t)$, the ground state fermion density $\rho_0(\mathbf{x})=\langle G|\hat{\rho}(\mathbf{x})|G\rangle $, $\hat{\rho}(\mathbf{x})=\hat{\Psi}^{\dag}(\mathbf{x})\hat{\Psi}(\mathbf{x})$, and a unit mass $m=1$ is taken. The $\chi^{\hat{\mathbf{j}}_p\hat{\mathbf{j}}_p}_{ij}(\mathbf{x},t;\mathbf{x}',t')$ is defined in the same way as $\chi^{\hat{\mathbf{j}}\hat{\mathbf{j}}}_{ij}(\mathbf{x},t;\mathbf{x}',t')$ in Eq.(\ref{retardedcorrfunctiontotalj}) but with $\hat{\mathbf{j}}^p$ in place of $\hat{\mathbf{j}}$.
Taking the time Fourier transform of Eq.(\ref{totalinducedcurrentStrinati}), we obtain 
\[
\delta j_i(\mathbf{x},\omega)=\int d\mathbf{x}'\chi^{\mathbf{j}_p\mathbf{j}_p}_{ij}(\mathbf{x},\mathbf{x}',\omega)A_j(\mathbf{x}',\omega)+\rho_0(\mathbf{x})A_i(\mathbf{x},\omega)
\]
\begin{equation}\label{totalinducedcurrentStrinatiFT}
=\int d\mathbf{x}'\left[\chi^{\mathbf{j}_p\mathbf{j}_p}_{ij}(\mathbf{x},\mathbf{x}',\omega)+\rho_0(\mathbf{x}')\delta(\mathbf{x}'-\mathbf{x})\delta_{ij}\right]A_j(\mathbf{x}',\omega)
\end{equation}
Comparing with Eq. (\ref{totalinducedcurrentFT}), we find that
\begin{equation}\label{magicrelation}
\chi^{\mathbf{j}\mathbf{j}}_{ij}(\mathbf{x},\mathbf{x}',\omega)=\chi^{\mathbf{j}_p\mathbf{j}_p}_{ij}(\mathbf{x},\mathbf{x}',\omega)+\rho_0(\mathbf{x}')\delta(\mathbf{x}'-\mathbf{x})\delta_{ij}
\end{equation}

Now consider the space-averaged response function that we propose; the dynamical current susceptibility. 
\begin{equation}
\chi^{\mathbf{j}\mathbf{j}}_{ij}(\omega)=\int d\mathbf{x}\int d\mathbf{x}'\chi^{\mathbf{j}\mathbf{j}}_{ij}(\mathbf{x},\mathbf{x}',\omega)
\end{equation}
Using Eq.(\ref{magicrelation}) we obtain
\begin{equation}\label{magicrelationfinal}
\chi^{\mathbf{j}\mathbf{j}}_{ij}(\omega)=\chi^{\mathbf{j}_p\mathbf{j}_p}_{ij}(\omega)+\delta_{ij}\int d\mathbf{x}\rho_0(\mathbf{x})
\end{equation}
where the last term is nothing but the ground state average number of electrons per unit cell $N_s=\int d\mathbf{x}\rho_0(\mathbf{x})$ for each spin species. This completes the derivation of Eq.(\ref{totaldyncurrsuscept}) in the main text.

\section{Consistency Check}
There are several consistency checks that can be done in order to verify the correctness of our current-based and linear response theory formalism.
It is first to be noted that we are working in linear response regime, which means the applied magnetic field $\mathbf{B}$ is weak and the response physical quantity; the polarization $\mathbf{p}$
is linear in $\mathbf{B}$. In this case, in the linear response Eq.(5) in the main text, the susceptibility tensor reflects a ground state property and the Hamiltonian from which it is computed does
not include the Zeeman magnetic field applied to the system. The axionic TME effect however enters via the vector potential $\mathbf{A}$, which has to be modified accordingly based on 
axion electrodynamics in order to describe such effect, as we noted earlier. 

\begin{figure}
 \includegraphics[scale=0.66]{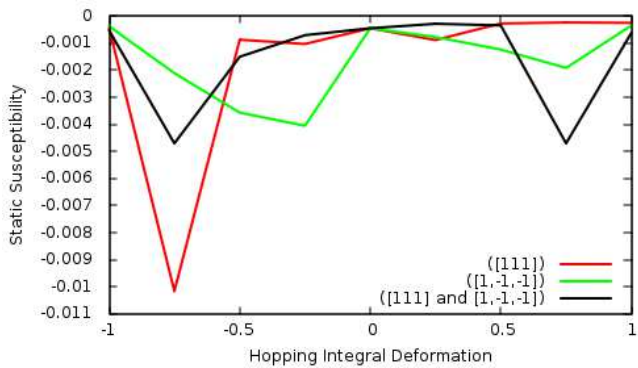}
 \label{fig:PvsGapNoSO}
 \caption{The static susceptibility $\chi^{\mathbf{j}_p\mathbf{j}_p}_{zz}(0)$ as function of hopping integral deformation $dt$ (proportional to band gap $\Delta=2|dt|$) along directions $[111]$,$[1,-1,-1]$, and both directions simultaneously, 
 in tight-binding model on diamond lattice without spin-orbit coupling. The units are such that $t=1.0$ and $a_l=1.0$ and we have used $\lambda_{SO}=0.125$.}
\end{figure}

As a consistency check, we present the result of our static susceptibility for model of tight-binding electron on diamond lattice without spin-orbit coupling. One expects to have normal metal for isotropic case $dt=0$ 
and ordinary insulator for any $0<|dt|\leq t$. 
Since the system is then ordinary insulator for both $dt>0$ and $dt<0$ with a change in the gap $\sim 2|dt|$ relative to that of the metal state at $dt=0$, the current susceptibility should behave the same way or become symmetric with respect to the critical gapless metal at $dt=0$ for hopping integral deformation along any direction. Since ordinary insulator is topologically trivial, clearly we expect the topological order parameter as we have defined to vanish all the way. That would confirm the consistency of the theory.

This is indeed what we obtain and confirm numerically as shown in Fig.3, where we use the same parameters as those used in obtaining Fig.1 in the main text, 
except that we set the spin-orbit coupling to infinitesimally small number, e.g. $\lambda_{\mathrm{SO}}=0.0001$.
We note that the magnitude of the static response without spin-orbit coupling is several orders of magnitude smaller than that of Fig.1 in the main text with spin-orbit coupling.
In addition, the former shows no regular pattern and this simply means it corresponds to numerical fluctuations. This simply suggests that this quantity vanishes in the 
zero spin-orbit coupling limit $\lambda_{\mathrm{SO}}=0$ (and thermodynamic limit). 
Using Eq.(\ref{OrderParameter}) in the main text, the order parameter $\delta \chi^s (dt)$ is thus always zero in the absence of spin-orbit coupling, as should indeed be the case. The 'residual' nonzero total current susceptibility $\chi^{\mathbf{j}\mathbf{j}}_{ij}(\omega)$, which from Eq.(\ref{totaldyncurrsuscept}) is simply a constant equals to the ground state average number of particles per unit cell per spin species for $i=j$ and zero otherwise, represents the ground state current correlation which does not come from the TME effect as the latter necessarily comes from the current induced by the applied magnetic field. As we know 3d TIs have surface state current even in the absence of external bias field and this contributes to  the ground state current that gives rise to nonzero ground state correlation.
This consistency further corroborates our identification of the dynamical current susceptibility $\chi^{\mathbf{j}\mathbf{j}}_{ij}(\omega)$ as a characterization of the TME effect and the order parameter $\delta \chi^s (dt)$ as an order parameter of the topological phase transition in 3d topological insulators.

\section{Details of Experimental Setup and Physical Explanation of Numerical Results on Dynamical Current Susceptibility}

The experiment that we propose in the main text can be illustrated as in Fig.4. A thick slab sample of 3d TI candidate material is coated on each of its surfaces with a layer of ferromagnet 
with magnetization $\mathbf{m}$ perpendicular to the interface \cite{QiHughesZhang}, whose dynamical spin wave (magnon) provides the dynamical
axion field $a$ that couples to the electric field $\mathbf{E}$ induced by a generally time-dependent spatially uniform magnetic field $\mathbf{B}$ applied parallel to the front and rear surfaces, which are normal to $[111]$ or $[1,-1,-1]$ direction. 
This magnetization is also needed to obtain the time reversal symmetry breaking in analogy to the effect of magnetic field in quantum Hall effect, in order to observe the TME effect \cite{QiHughesZhang}.

Consider a static magnetic field first. 
Then there will be a polarization parallel to  the magnetic field and which induces
a charge density of opposite signs at the bottom and top ends of the sample. Since the axion field is dynamical, this polarization is readily time-dependent and gives rise to a current, following from
$\mathbf{j}=\partial_t\mathbf{p}$ and a time-dependent charge density according to continuity equation $\nabla\cdot\mathbf{j}+\partial_t\rho=0$.  
Applying time-dependent magnetic field gives rise to additional current which together with the charge density always satisfies the continuity equation.  In this setup, the current will be uniformly distributed over the front and rear surfaces (assuming edge effect is negligible) so that point contact measurement is equivalent anywhere on the surface, corresponding therefore directly to the space-averaged value of the quantity of interest.

\begin{figure}
	\includegraphics[scale=0.40]{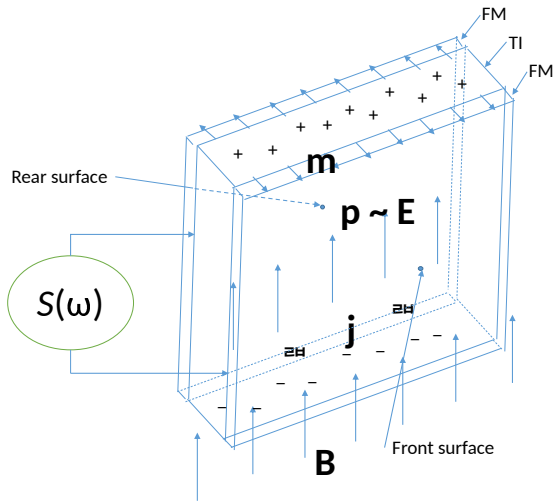}
	\label{fig:Drawings}
	\caption{Experimental setup for measurement of current noise correlation from topological magnetoelectric effect. The $S(\omega)$ measures the correlation between the currents on the front and rear surfaces of the TI \cite{Notesidesurface}. }
\end{figure}

Now suppose that the current is slowly increasing on the front surface, then the charge density on the bottom front as a result is decreasing. But charge conservation 
demands that the current on the rear surface should be decreasing and the charge density at the rear bottom end be slowly increasing, for slow enough time variation of the current. 
This implies that the currents on the front and rear surfaces are anticorrelated
at small frequencies. This is in complete agreement with our explicit calculations represented in Fig. 1 and Fig. 2 where the current cross-correlation is negative at DC and low frequencies $0\leq \omega < \omega_0$.

At larger than critical frequencies, the currents on the two surfaces are positively correlated and there will be flow of charge from inside the bulk to supply for charge accumulation or 
depletion at the top and bottom ends of the sample. This is easy to understand if we note that in the energy band dispersion in $\mathbf{k}$ space, low frequency photons only excite surface state electrons 
or states near Fermi energy whereas high frequency photons excite electrons deep within bulk bands. The explanation for why the strength of sign-changing singularity is weaker in the (strong) topological insulating than in the weak topological (or ordinary) insulating phase is more curious but intuitively it relates to the surface state. In general, the presence of surface state facilitates the flow of charge and helps the electromagnetic field to excite low energy electrons near Fermi energy. 
Since the surface state of STI encloses an odd number of Dirac points and so has at least one such Dirac point, the transition from negative to positive current correlation is smoother as it is facilitated by the surface state, compared to a WTI or ordinary insulator, which has no surface state or has surface state that encloses an even number of Dirac points (the effect of which may cancel out each other), the photons work by themselves in exciting the electrons without extra assistance from the surface state and this results in more dramatic sudden change from the negative to positive current correlation across the critical frequency $\omega_0$.


\begin{thebibliography}{1}

\bibitem{K-M}C.L. Kane and E.J. Mele, Phys. Rev. Lett. 95, 226801 (2005); Phys. Rev. Lett. 95, 146802 (2005).

\bibitem{BernevigZHang}Bernevig, B. A., T. A. Hughes, and S. C. Zhang, Science 314, 1757 (2006); B. A. Bernevig and S. C. Zhang, Phys. Rev. Lett. 96, 106802 (2006).

\bibitem{ZHasanKaneRMP}M. Z. Hasan and C. L. Kane, Rev. Mod. Phys. 82, 3045 (2010).

\bibitem{FKM}L. Fu, C. L. Kane, and E. J. Mele, Phys. Rev. Lett. 98, 106803 (2007).

\bibitem{QiZhangRMP}X.-L. Qi and S.-C. Zhang, Rev. Mod. Phys. 83, 1057 (2011).

\bibitem{QiHughesZhang}X-L. Qi, T. L. Hughes, and S-C. Zhang, Phys. Rev. B 78, 195424 (2008).

\bibitem{WilczekAXION}F. Wilczek, Phys. Rev. Lett. 58, 1799 (1987).

\bibitem{RosenbergFranz}G. Rosenberg and M. Franz, Phys. Rev. B 82, 035105 (2010).

\bibitem{dynamicalaxion}R. Li, J. Wang, X.-L. Qi and S.-C. Zhang, Nature Physics 6, 284
(2010).

\bibitem{Balents-Moore-RahulRoy}J. E. Moore and L. Balents, Phys. Rev. B 75, 121306(R)(2007),

\bibitem{Balents-Moore-RahulRoy2}R. Roy, Phys. Rev. B 79, 195322 (2009).

\bibitem{Fu-KaneTIinversionsymmetry}L. Fu and C. L. Kane, Phys. Rev. B 76, 045302 (2007).

\bibitem{MurakamiNJP}S. Murakami, New J. Phys. 9, 356 (2007).

\bibitem{KingSmithVanderbilt}R. D. King-Smith and D. Vanderbilt, Phys. Rev. B 47, 1651 (1993).

\bibitem{OrtizMartin}G. Ortiz and R. M. Martin, Phys. Rev. B 49, 14202 (1994).

\bibitem{RestaReview}R. Resta, J.Phys.: Condens.Matter 22 (2010) 123201.

\bibitem{Strinati}G. Strinati, Rivista del Nuovo Cimento 11, 12 (1988).

\bibitem{PRthesis}P. Romaniello, Time-Dependent Current-Density-Functional Theory for Metals, PhD Thesis (2006).

\bibitem{orbital}We only consider the orbital current in this work.

\bibitem{orbitaldirac}Again, we retain only the orbital part of the current in the relativistic case.

\bibitem{PRprl}N. Raimbault, P. L. de Boeij, P. Romaniello, and J. A. Berger, Phys. Rev. Lett. 114, 066404 (2015).

\bibitem{SuppMat}Please see the Appendices.

\bibitem{kineticmassaxion}The axion kinetic and mass terms $(\partial_{\mu}a)^2/2-m^2a^2/2$ which ensure that $\langle a \rangle \neq 0$ in general, do not affect the field equations.

\bibitem{Notesinderivation}We have assumed spatially uniform field $a(\mathbf{r},t)=a(t)$.Extension to general $a(\mathbf{r},t)$ is straightforward though tedious.

\bibitem{Notesassumption}Without loss of generality, we have taken $\mathbf{A}(\mathbf{r}, t=-\infty)=0$ and $\mathbf{E}_0(\mathbf{r})=\mathbf{E}(\mathbf{r},t=-\infty)=0$.

\bibitem{Chemistry}J. Van Der Rest and P. Pecheur, Journal de Physique, 1983, 44 (11), pp.1297-1305.

\bibitem{Noteimag}The imaginary part in Fig.1 is zero while in Fig.2 is Dirac delta function centered at the critical frequency.

\bibitem{Exps} S.-Y. Xu, Y. Xia, L. A. Wray, S. Jia, F. Meier, J. H. Dil,
J. Osterwalder, B. Slomski, A. Bansil, H. Lin, R. J. Cava, and M. Z. Hasan, Science 332, 560 (2011); T. Sato, K. Segawa, K. Kosaka, S. Souma, K. Nakayama, K.  Eto,  T.  Minami,  Y.  Ando,  and  T.  Takahashi,  Nat. Phys. 7, 840 (2011); L.  Wu,  M.  Brahlek,  R.  V.  Aguilar,  A.  V.  Stier,  C.  M.
Morris, Y. Lubashevsky, L. S. Bilbro, N. Bansal, S. Oh, N. P. Armitage, Nat. Phys. 9, 410 (2013). 

\bibitem{Vanderbilt}A. M. Essin, J. E. Moore, and D. Vanderbilt, Phys. Rev. Lett. 102, 146805 (2009). 

\bibitem{Wenbook}Wen, X.-G., 2004, Quantum Field Theory of Many-Body Systems (Oxford University Press, Oxford).

\bibitem{QHtransition}The Quantum Hall Effect, Ed. by R. E. Prange and Steven M. Girvin (Springer, 1990).

\bibitem{Imry}Y. Imry, Introduction to Mesoscopic Physics (Oxford University Press (1997)). 

\bibitem{axioncoldatom}A. Bermudez, L. Mazza, M. Rizzi, N. Goldman, M. Lewenstein, and M.A.
Martin-Delgado, Phys. Rev. Lett. 105, 190404 (2010).

\bibitem{GiulianiS}G. Giuliani and G. Vignale, Quantum Theory of the Electron Liquid (Cambridge University Press (2015)).

\bibitem{Notes}The axion field $a$ can in general be function of space and time but for the response of topological insulators, topological field theory \cite{QiHughesZhang} 
	shows that the TME effect comes from the time dependence of the axion field $a$.
	
\bibitem{Notesidesurface}The side surfaces of the TI should also be coated with ferromagnetic layer (not shown in the figure) but in such a way that still allows the measurement of the currents on the front and rear surfaces.


\end{thebibliography}
\end{document}